\documentclass{elsart}

\usepackage{amssymb,amsmath,mathrsfs,eufrak}

\newtheorem{exm}{Example}[section]

\newcommand{\A}{\mathcal{A}}
\newcommand{\B}{\mathcal{B}}

\newcommand{\N}{\mathbf N }

\newcommand{\nl}{\\ \hspace*{0.3em} }

\newcommand{\ar}{\rightarrow}
\newcommand{\la}[1]{\it {#1}\rm}

\newcommand{\Hy}{\mathcal{H}}
\newcommand{\T}{ {\mathbf T \rm} }

\newcommand{\mL}{\mathcal{L}}

\newcommand{\E}{ {\mathcal E} }

\newcommand{\C}{{\mathbf C}}

\newcommand{\1}{{\bf 1}}
\newcommand{\mN}{\mathbb{N}}

\newcommand{\mF}{\mathbb{F}}

\newcommand{\Set}{ {\bf Set} }
\newcommand{\F}{\mathcal{F}}

\newcommand{\ld}{\boldsymbol{\lambda}}

\newcommand{\lf}{\boldsymbol{\forall}}

\newcommand{\lx}{\boldsymbol{\exists}}

\newcommand{\diam}{\diamond}
\newcommand{\df}[1]{\begin{defn} #1 \end{defn}}

\newcommand{\lm}[1]{\begin{lem} #1 \end{lem}}
\newcommand{\te}[1]{\begin{thm} #1 \end{thm}}
\newcommand{\ex}[1]{\begin{exm} #1 \end{exm}}

\newcommand{\rk}[1]{\begin{rem} #1 \end{rem}}
\newcommand{\ps}[1]{\begin{prop} #1 \end{prop}}

\newcommand{\TT}{{\mathsf T }}
\newcommand{\FF}{{\mathsf F }}

\newcommand{\tb}{{\tilde b }}
\newcommand{\ta}{{\tilde a }}
\newcommand{\tc}{{\tilde c }}
\newcommand{\td}{{\tilde d }}
\newcommand{\tx}{{\tilde x }}
\newcommand{\tm}{{\tilde m }}

\newcommand{\tee}{{\tilde e }}

\newcommand{\bF}{{\mathbb F}}
\newcommand{\bR}{{\mathbb R}}

\begin{document}

\begin{frontmatter}



\title{\LARGE{Clone Theory} \\ \Large{Its Syntax and Semantics}\\
 \large{Applications to \\ Universal Algebra, Lambda Calculus \\ and Algebraic Logic} }


\author{Zhaohua Luo}

\small{July 2007

(Revised October 2008)}

 \ead{zluo@azd.com}

\begin{abstract}

The primary goal of this paper is to present a unified way to
transform the syntax of a logic system into certain initial
algebraic structure so that it can be studied algebraically. The
algebraic structures which one may choose for this purpose are
various clones over a full subcategory of a category. We show that
the syntax of equational logic, lambda calculus and first order
logic can be represented as clones or right algebras of clones over
the set of positive integers. The semantics is then represented by
structures derived from left algebras of these clones.

\end{abstract}

\end{frontmatter}


\section*{Introduction}

Let $\mN = \{1, 2, 3, ...\}$ be the set of positive integers. Denote
by $\Set$ the category of sets.

We will introduce the following fundamental structures for universal
algebra, lambda calculus and first order logic:
\\1. Clones over $\mN$.
\\2. Left and right algebras of a clone over $\mN$.
\\3. $\lambda$-clones, $\lambda_{\beta}$-clones (reflexive clones).
\\4. $\lambda$-algebras.
\\5. Predicate algebras with terms in a clone over $\mN$.
\\6. Quantifier algebras with terms in a clone over $\mN$.

Note that the class of objects in each category is a variety in the
sense of universal algebra. Four basic observations on
algebraization are:
\\(i) Finitary endofunctors of $\Set$ are represented by locally
finitary right algebras of the initial clone over $\mN$.
\\(ii) Finitary monads in $\Set$ (or finitary varieties) are represented by locally finitary clones
over $\mN$.
\\(iii) The set of $\lambda$-terms is represented by the initial
$\lambda$-clone.
\\(iv) The set of formulas of a first order language is represented by a
predicate algebra over the clone of terms.

The theory of clones considered in this paper originated from the
theory of monads. Two equivalent definitions of monads, namely
\la{monads in clone form} and \la{monads in extension form} given by
E. Mane~\cite{mane}, can be interpreted as only defined over a given
subcategory of the category. These are \la{clones} and \la{clones in
extension form over a full subcategory}. It turns out that these two
new concepts are no longer equivalent unless the subcategory is
dense. But morphisms of clones, algebras of clones, and morphisms of
algebras can all be defined for these two types of clones. Since
many familiar algebraic structures, such as monoids, unitary Menger
algebras, Lawvere theories, countable Lawvere theories, classical
and abstract clones are all special cases of clones over various
dense subcategories of $\Set$, the syntax and semantics of these
algebraic structures can be developed in a unified way, so that it
is much easier to extend these results to many-sorted sets.

Just as monads arise from adjunctions of categories, clones arise
from ``adjunctions'' of \la{species}. The notions of clones and
species introduced in this paper are very easy to manipulate, yet
they are more flexible than the traditional notions of monads and
adjunctions.

Whenever it is convenient in this paper composition of morphisms in
a category is written from left to right.

A \la{species} $\N/\C$ consists of a category $\C$ and a full
subcategory $\N$ of $\C$; if $\N = \C$ then we say that $\N/\N$ (or
simply $\N$) is a \la{singular species}. If $\N'/\C'$ is another
species, a \la{function} (resp. \la{functor}) $T$ from $\N'/\C'$ to
$\N/\C$ is a function $T: Ob \C' \ar Ob \C$ (resp. functor $T: \C'
\ar \C$) such that
\\(i). $Ob \N' = Ob \N$.
\\(ii). $\C'(A, B) = \C(A, TB)$ for $A \in \N$ and $B \in \C'$.
\\(iii).  $r(fg) = (rf)g$ (resp. $f(Tg) = fg$) for $C, D \in \N$, $r \in \N(D,
C)$, $f \in \C'(C, A)$ and $g \in \C'(A, B)$. \\Note that the
composite of two functions (or functors) of species is a function
(or functor) of species.

A \la{clone theory} (resp. \la{clone theory in extension form}) over
a full subcategory $\N$ of a category $\C$ is a pair $(\N', T)$
where $\N'$ is a category and $T$ is a function (resp. functor) of
species from $\N'/\N'$ to $\N/\C$. We often simply say that $\N'$ or
$T$ is a clone theory over $\N$.

Suppose $\N/\C$ is a species. A \la{species over} $\N/\C$ is a pair
$(\N'/\C', T)$  consists of a species $\N'/\C'$ and a functor $T:
\N'/\C' \ar \N/\C$.

If $(\N'/\C', T)$ is a species over $\N/\C$ then $(\N', T|_{\N'})$
is a clone theory over $\N$, called the \la{clone theory of
$\N'/\C'$}, denoted by $Clone((\N'/\C'), T)$. Note that $\N'$
consists of free objects of $\C'$ over $\N$ with respect to $T: \C'
\ar \C$.

Let $(\N', T)$ be a clone theory in extension form over $\N$. By an
\la{$(\N', T)$-species} we mean a species over $\N/\C$ with $(\N',
T)$ as its clone theory.

The class $Sp(\N', T)$ of $(\N', T)$-species viewed as concrete
categories over $\C$ forms a meta-category. Clearly $(\N', T)$ is
the initial object of $Sp(\N', T)$. $Sp(\N', T)$ has a canonical
terminal object $\C^T$, called the Eilenberg-Moore species of $(\N',
T)$, consisting of $T$-algebras (see Definition~\ref{algebra}).

\rk{1. A functor from a singular species to a species is a clone
theory in extension form.

2. A functor from a species to a singular species is equivalent to
an adjunction of categories.

3. A functor from a singular species to a singular species is
equivalent to a monad.

4. Any functor from a species to another species determines a clones
theory in extension form. }

If $\N$ is dense in $\C$ then the notion of a clone theory over $\N$
is equivalent to that of a clone theory over $\N$ in extension form.
In practice one is primarily interested in the clone theories over a
dense subcategory. For this reason if $\N$ is not dense in $\C$ we
always choose a proper subcategory $\C'$ of $\C$ containing $\N$
such that $\N$ is dense in $\C'$. Hence the distinction between the
two types of clone theories is not essential.

Since a clone theory (or clone theory in extension form) over $\N$
gets objects and morphisms from $\C$, they can be intrinsically
defined as algebraic systems in $\C$, called \la{clone systems}. The
study of clone systems is presented in Section 1. Since clone theory
and clone system are essentially the same, they are often referred
just as a clone.

If $(\N', T)$ is a clone over $\N$, then there is a functor $F^T: \N
\ar \N'$ preserving objects and sending each $f: A \ar B$ in $\N$ to
$f\eta: A \ar TB$ in $\N'$, where $\eta: B \ar TB$ is the identity
of $B \in \N'$. Linton in~\cite{linton:2} defines a more general
notion of a clone over a subcategory $\N$  of a category as an
arbitrary functor $\N \ar \N'$ which is bijective on objects. Our
definition of algebras for a clone theory coincides with that of
Linton's under functorial consideration. Thus all the results
of~\cite{linton:2} apply to clone theories. We mention that all the
examples given in~\cite{linton:2} p.22 are in fact clones over dense
subcategories.

\ex{Examples of clones over dense subcategories are abundant. Here
are some of the most elementary examples.
\\1. A clone over a singleton in $\Set$ is a monoid.
\\2. A clone over a nonempty set $N$ in $\Set$ is a unitary Menger
algebra $T$ of rank $|N|$.
\\3. A clone (resp. the dual of the clone theory) over the full
subcategory $\{0, 1, 2, 3, ...\}$ of $\Set$ is a clone in the
classical sense (resp. Lawvere theory)  (note that here each integer
$n \ge 0$ is viewed as a finite set with $n$ elements).
\\4. The dual of a clone theory over the full subcategory $\{0,
1, 2, 3, ..., \mN\}$ (or $(0, 1, \mN)$) of $\Set$ is a countable
Lawvere theory in the sense of \cite{po:1}.
\\5. The dual of a clone theory over the full subcategory $\{1,
2, 3, ..., \mN\}$ (or $(1, \mN)$) is equivalent to an algebraic
theory (see section~\ref{sec:at}).
\\6. A clone over a one-object-category is a Kleisli algebra. Any
clone over an infinite set $N$ in $\Set$ defines a Kleisli algebra
(see section~\ref{sec-lc}).
\\7. Other important examples are clones over the subcategory of
finitely presentable objects of a locally finitely presentable
category.
\\8. A clone (or clone in extension form) over $\N = \C$
is equivalent to a monad in extension form, or a Kleisli triple in
$\C$ in the usual sense. Hence the notion of a clone generalizes
that of a monad. }

The simplest type of clones are clones over an object $N$ of a
category $\C$. Recall that a \la{left act of a monoid $M$} (or a
\la{left $M$-act}) is a set $D$ together with a map $M \times D \ar
D$ such that $ed = d$ where $e$ is the unit of $M$ and $m(m'd) =
(mm')d$ for any $m, m' \in M$ and $d \in D$. A \la{right act of $M$}
is defined similarly.

\rk{A clone over an object $N$ of a category $\C$ is an object $\A$
of $\C$ such that $hom(N, \A)$ is a monoid and $r(fg) = (rf)g$ for
all $r: N \ar N$ and $f, g: N \ar \A$. Suppose $\A$ is a clone over
$N$. A left $\A$-algebra is an object $D$ such that $hom(N, D)$ is a
left act of the monoid $hom(N, \A)$ and $r(fm) = (rf)m$ for all $r:
N \ar N$, $f : N \ar \A$ and $g: N \ar D$. A right $\A$-algebra is a
right act of $hom(N, \A)$.}

\rk{ A clone in extension form over an object $N$ of a category $\C$
is an object $\A$ such that $hom(N, \A)$ is a monoid together with a
homomorphism $T: hom(N, \A) \ar hom(\A, \A)$ of monoids such that
$f(Tg) = fg$ for $f,g: N \ar \A$. A left $\A$-algebra is then an
object $D$ such that $hom(N, D)$ is a left act of the monoid $hom(N,
\A)$ together with a homomorphism $H: hom(N, D) \ar hom(\A, D)$ of
left acts of $hom(N, \A)$ such that $f(Hg) = fg$ for $f: N \ar \A$
and $g: N \ar D$. A right $\A$-algebra is a right act of $hom(N,
\A)$. }

Assume $\C = \Set^S$ for a set of sorts $S$, and $N = \{N_s\}_{s \in
S}$ is an $S$-sorted set. Then a clone in extension form over $N$ is
an $S$-sorted set $\A = \{\A_s\}_{s \in S}$ such that $hom(N, \A)$
is a monoid and there is a homomorphism $T: hom(N, \A) \ar hom(\A,
\A)$ of monoids such that $f(Tg) = fg$. Since \[hom(\A, \A) =
\prod_{s\in S} hom(\A_s, \A_s),\] $T$ is uniquely determined by a
sequence of maps $\A_s \times hom(N, \A) \ar \A_s$. Thus
algebraically a clone in extension form over $N$ can be defined as
follows:

\begin{defn}\label{def:clone}
A clone in extension form over an $S$-sorted set $N$ is an
$S$-sorted set $\A$ together with maps $\{\mu_s: \A_s \times hom(N,
\A) \ar \A_s\}_{s \in S}$ and a map $x: N \ar \A$ such that for any
$a \in \A_s$ and $f = \{f_s\}, g = \{g_s\} \in hom(N, \A)$:
\\(i) $(af)g = a(fg)$ where $(fg)_{si} = f_{si}g$ for any $s \in S$, $i \in N_s$ and $f_{si} = f_s(i)$.
\\(ii) $x_{si}f = f_{si}$.
\\(iii) $ax = a$.
\end{defn}

If $\A$ is a clone in extension form over an $S$-sorted set $N$ then
each $\A_s$ is a right $hom(N, \A)$-act and $hom(N, \A)$ is the
product $\prod_{s\in S} hom(N_s, \A_s)$ of these right $hom(N,
\A)$-acts. Categorically a clone in extension form over $N$ is a
category with a dense object $\A^*$ and an $S$-indexed set of
objects $\{\A_s\}_{s \in S}$ such that $\A^*$ is the product
$\prod_{s\in S} \A_s^{N_s}$.

Let $\Set^S_*$ be the set $N$ of $S$-sorted sets such that $N_s$ is
not empty for any $s \in S$. An $S$-sorted set $N$ is dense in
$\Set^S_*$ if each $N_s$ has at least two elements, thus the notions
of a clone and a clone in extension form over such $N$ are
equivalent.

Let $\mN$ be the set of positive integers. Let $\mN_S =
\{\mN\}_{s\in S}$. Then the above analysis applies to clones over
$\mN_S$. Let $\A$ be a clone over $\mN_S$. A left $\A$-algebra has
the following form:

\df{A left $\A$-algebra is an $S$-sorted set $D$ together with maps
$\{\mu_s: \A_s \times D^{\mN} \ar D_s\}_{s\in S}$ such that for any
$a \in \A_s$, $f \in \A^{\mN}$ and $g \in D^{\mN}$:
\\(i) $(af)g = a(fg)$ where $(fg)_{si} = f_{si}g$ for any $s \in S$ and $i \in N_s$.
\\(ii) $x_{si}g = g_{si}$.}

Denote by $Rg(\A)$ and $Lg(\A)$ the categories of right and left
$\A$-algebras respectively. Then $Rg(\A)$ is a topos and $Lg(\A)$ is
an algebraic category over $\Set^S$.

A clone $\A$ over $\mN_S$ is \la{locally finitary} if for any $a \in
\A_s$ we have $ae = a$ for a map $e: \mN_S \ar \A$ such that
$e(\mN_S) \subseteq x(\mN_S)$, $e(\mN_S)$ has finite components and
$ee = e$; $e(\mN_S)$ is then called a \la{support of} $a$. The
assertions 2 and 3 of the following main theorem for many-sorted
clones extend a similar theorem for (one-sorted) clones over $\mN$
due to W. D. Neumann~\cite{newmann:rep}:

\te{1. The full subcategory of locally finitary clones over $\mN_S$
is a coreflective subcategory of the category of clones over
$\mN_S$. \\2. The class of left $\A$-algebras of a locally finitary
clone $\A$ over $\mN_S$ is a finitary $S$-sorted variety. \\3.
Conversely, any finitary $S$-sorted variety $V$ as a concrete
category over $\Set^S$ is equivalent to the category of left
algebras of the clone determined by the free algebra of $V$ on
$\mN_S$. }

Suppose $B$ is any right $\A$-algebra. In preparation for the
definitions of $\lambda$-clones and predicate algebras with terms in
$\A$ we need an explicit form for the exponent $B^{\A_s}$ in the
cartesian closed category $Rg(\A)$ with respect to the right
$\A$-algebra $\A_s$. It is a crucial fact for our purpose that the
underlying set of $B^{\A_s}$ can be identified with the set $B$
itself, so that a homomorphism $B^{\A_s} \ar B$ reduces to a unary
operation $B \ar B$ (which is not an endomorphism of right
$\A$-algebras in general). This can be seen as follows.

Since $\A_s$ and $B$ are two right acts of the monoid $\A^{\mN_S}$,
the right act $B^{\A_s}$ can be defined as $\hom(\A^{\mN_S} \times
\A_s, B)$ with action $f \ar fu$ of $u: \mN_S \ar \A$ on $f:
\A^{\mN_S} \times \A_s \ar B$ defined by $(fu)(u', b) = f(uu', b)$
for any $u': \mN_S \ar \A$ and $b \in B$ (cf.~\cite{maclane}, p.62,
ex.5). Since $\mN$ is infinite, $\A_s^{\mN} \times \A_s$ is
isomorphic to $\A_s^{\mN}$ as right $\A$-algebras. Since
$\A_s^{\mN}$ is generated by $x_s = [x_{s1}, x_{s2}, ...]$,
$\A_s^{\mN} \times \A_s$ is generated by $x^*_{s1} = ([x_{s2},
x_{s3}, ...], x_{s1})$. Fixed such an isomorphism via $x^*_{s1}$. We
obtain an isomorphism $\A^{\mN_S} \times \A_s = \prod_{s \in S}
\A_s^{\mN_S} \times \A_s$ to $\A^{\mN_S} = \prod_{s \in
S}\A_s^{\mN_S} $. Since $\A^{\mN_S}$ is the free right act generated
by the unit $x$, there is a bijective map $\hom(\A^{\mN_S}, B) \ar
B$ which maps $f$ to $f(x)$ for $f: \A^{\mN_S} \ar B$. Thus the
underlying sets of the following right $\A^{\mN_S}$-acts are
bijective
\[B^{\A_s} = \hom(\A^{\mN_S} \times \A_s, B) \cong \hom(\A^{\mN_S}, B) \cong B.\]
Note that as right $\A$-algebra $B^{\A_s}$ and $B$ are not
isomorphic in general. In section~\ref{sec-lc} we will give a direct
definition of $B^{\A_s}$ using $B$ (for the one-sorted case). The
extension for many-sorted cases is straightforward.

Formally one may just specify a type of unary operations on $B$
corresponding to homomorphisms from $B^{\A_s}$ to $B$ (under the
identification of $B^{\A_s}$ with $B$ via $x_{s1}^*$), called \la{an
abstract binding operation (on $x_{s1})$}. Unlike classical binding
operation which binds a specific variable, an abstract binding
operation only reduces the size of the support of a finitary element
of $B$ by $0$ or $1$. Elementary properties of abstract binding
operations (for the one-sorted case) are given in
section~\ref{sec-bo}. The importance of an abstract binding
operations lies in the fact that the classical unary binding
operations such as
\[\ld x_1, \ld x_2, \ld x_3, \ld x_4,...\]
\[\lf_s x_1, \lf_s x_2, \lf_s x_3, \lf_s x_4,... (s \in S)\]
 \[\lx_s x_1, \lx_s x_2, \lx_s x_3, \lx_s x_4,...(s \in S)\] can be defined
as derived operations from abstract binding operations: $\lambda$,
$\forall_s$ and $\exists_s$, respectively, provided the
substitutions of variables have been defined. Thus to define
$\lambda$-terms or formulas for a first order language, one could
take abstract binding operations (e.g. $\lambda$ or $\forall_s$) as
the basic operation, and define the other classical binding
operations via substitutions, as $\lambda$-terms form a clone, and
formulas of a first order language form a right algebra of the clone
of terms. This is the ideal approach to solve the problems of
$``$variable capture$"$ caused by substitutions in lambda and
predicate calculus: at one hand the troublesome process of renaming
bound variables is push into the proper background, on the other
hand the classical binding operations are still available, which are
more readable than an abstract binding operations. This is probably
well known for lambda calculus (cf. P. Aczel~\cite{aczel:1} 2.3.1).

\df{A \la{$\lambda$-clone} is a clone $\A$ over $\mN$ together with
two homomorphisms $\A^2 \ar \A$ and $\A^{\A} \ar \A$ of right
$\A$-algebras.} The class of $\lambda$-clones forms a (non-finitary)
variety. Therefore free $\lambda$-clones over any give set (called
``holes'' in literature) exist. The initial $\lambda$-clone is
precisely the set of $\lambda$-terms in de Bruijn's notation (or the
set of classical $\lambda$-terms modulo $\alpha$-conversions). A
clone $\A$ over $\mN$ is called \la{reflexive} if $\A^\A$ is a
retract of $\A$, \la{extensional} if $\A^{\A}$ is isomorphic to
$\A$. One can  show that a clone is reflexive (resp. extensional) if
and only if it satisfies the axiom for $\beta$-conversion (resp.
both $\beta$ and $\eta$-conversions). The class of reflexive clones
also forms a variety. The left algebras of the initial reflexive
clone are \la{$\lambda$-algebras} considered in
literature(cf.~\cite{ba:1}).

\df{An \la{$S$-sorted predicate algebra with terms in an $S$-sorted
clone $\A$ over $\mN_S$} is a right $\A$-algebra $P$ with the
following homomorphisms:
\\1. $\Rightarrow: P^2 \ar P$,
\\2. $\FF: P^0 \ar P$.
\\3. $\forall_s: P^{\A_s} \ar P$ for each $s \in S$.
\\We also assume that there are identities $\approx_s \ \in P$ for each $s
\in S$, where $\approx_s$ has support $\{x_{s1}, x_{x2}\}$.} Any
left $\A$-algebra $D$ determines a predicate algebra $P(D^{\mN_S})$
where $P(D^{\mN_S})$ is the power set of $D^{\mN_S}$. A \la{model of
a predicate algebra} $P$ is then a pair $(D^{\mN_S}, \mu)$ where $D$
is a left $\A$-algebra and $\mu: P \ar P(D^{\mN_S})$ is a
homomorphism of predicate algebras. An element $p \in P$ is
\la{logic valid} if $\mu(p) = \TT$ for any model of $P$, where $\TT
= (\FF \Rightarrow \FF)$. For an $S$-sorted first-order language
$\mL$ the set $T(\mL)$ of terms of $\mL$ is naturally a clone over
$\mN_S$, and the set $F(\mL)$ of formulas of $\mL$ is a right
$T(\mL)$-algebra. which is an $S$-sorted predicate algebra with
terms in $T(\mL)$. The proof theory and model theory of $\mL$ can
then be carried out by algebraic considerations for the predicate
algebra $F(\mL)$ and its models.

The paper is organized as follows. Section 1 consists of formal
definitions of clones and left algebras of clones over a
subcategory. In section 2 we study various properties of clones over
$\mN$. As applications of clone theory to universal algebra we
discuss briefly how to define the notions of hyperidentities and
hypervarieties in terms of $C$-clones. Also a purely algebraic
approach to Morita theory for finitary varieties (i.e. locally
finitary clones) is sketched at the end of the section. In section 3
we introduce a new type of algebraic theories consisting of only two
objects, which are equivalent to clones over $\mN$. In section 4 a
general theory of binding unary operations on a right algebra of a
clone is introduced. It is used to define $\lambda$-clones in
section 5. Section 6 contains a simple modified classical approach
to one-sorted first-order theory, using De Bruijin style formulas to
provide a better substitution theory. The notion of one-sorted
predicate algebra is given in section 7.

\section{Clones
}\label{sec-clones}

Let $\C$ be a category. Let $\N$ be a full subcategory of $\C$.

\df{A clone over $\N$ is a system $T = (T, \eta, *)$ consisting of
functions
\\ (a) $T: Ob \N \ar Ob \C$,
\\(b) $\eta$ assigns to each object $A$ in $\N$ a morphism $\eta_A:
A\ar TA$,
\\(c) $*$  assigns to each ordered triple $(A, B, C)$  of objects of
$\N$ a function \[*: \C(A, TB) \times \C(B, TC) \ar \C(A, TC)\] such
that for any $r: A \ar B$, $h: A \ar TB$, $f: B \ar TC$ and $g: C
\ar TD$ with $D \in \N$ we have
\\(i) $(h * f) * g = h * (f * g)$.
\\(ii) $ (r\eta_B)* f   = rf$.
\\(iii) $ f * \eta_C  = f$.
\\Let $T = (T, \eta, *)$ and $T' = (T', \eta, \circ)$ be two
clones over $\N$. For each object $A$ of $\N$ let $\rho_A: TA \ar
T'A$ be a morphism. Then $\rho$ is a morphism of clones over $\N$ if
it preserves $\eta$ and $(f * g)\rho_C = (f \rho_B)\circ(g \rho_C)$
for any $f: A \ar TB$ and $g: B \ar TC$.}

\df{Suppose $T$ is a clone over $\N$. A left $T$-algebra is a pair
$X = (X, \circ)$ consisting of an object $X$ of $\C$ and a function
$\circ$ which assigns to each ordered pair $(A, B)$ of objects of
$\N$ a function $\C(A, TB) \times \C(B, X) \ar \C(A, X)$ such that
for any $k: A \ar C$, $f: A \ar TB$, $g: B \ar TC$ and $h: C \ar X$
with $D \in \N$ we have
\\(i) $(f \circ g) \circ h = f \circ (g \circ h)$.
\\(ii) $ (k\eta_C)\circ h   = kh$.
\\Let $X$ and $Y$ be two $T$-algebras. A morphism of left
$T$-algebras from $X$ to $Y$ is a morphism $\phi: X \ar Y$ such that
$(f \circ m)\phi = f \circ (m \phi)$ for any $f: A \ar TB$ and $m: B
\ar X$. }

\df{A clone over $\N$ in extension form is a system $T = (T, \eta,
*-)$ consisting of functions
\\(a) $T: Ob \N \ar Ob \C$,
\\(b) $\eta$ assigns to each object $A$ in $\N$ a morphism $\eta_A: A
\ar TA$,
\\(c) $*-$  maps each morphism $f: B \ar TC$ with $B, C$ in $\N$ to
a morphism $*f: TB \ar TC$, such that for $g: C \ar TD$ with $D \in
\N$
\\(i) $*f* g  = *(f* g )$.
\\(ii) $ \eta_B*f  = f$.
\\(iii) $ *\eta_C   = id_{TC}$.
\\Let $T$ and $T'$ be two clones in extension form over $\N$. For
each object $A$ of $\N$ let $\rho_A: TA \ar T'A$ be a morphism. Then
$\rho$ is a morphism of clones if it preserves $\eta$ and $*f \rho_B
= \rho_A *(f \rho_B)$ for any $f: A \ar TB$ with $B \in \N$.}

\df{\label{algebra} Suppose $T$ is a clone in extension form over
$\N$. A left $T$-algebra is a pair $X = (X, *-)$ consisting of an
object $X$ of $\C$ and a function $*-$ which maps each morphism $m:
A \ar X$ ($A \in \N$) to a morphism $*m: TA \ar X$, such that
\\(i) $ (*g)(*m) = *(g*m)$ for any $g: B \ar TA$ with $B \in \N$.
\\(ii) $\eta_X *m  = m$.
\\Let $X$ and $Y$ be two left $T$-algebras. A morphism of left
$T$-algebras from $X$ to $Y$ is a morphism $\phi: X \ar Y$ such that
$(*m)\phi = *(m \phi)$ for any $m: A \ar X$ with $A \in \N$. }

The class $\C^T$ of left $T$-algebras of a clone $T$ in extension
form over $\N$ is a concrete category over $\C$ with the forgetful
functor $G^T: \C^T \ar \C$. On the other direction we have a free
functor $F^T: \N \ar \C^T$ defined by $F^T(A) = (TA, *-)$ and
$F^T(f) = f\eta_A$. If $\N = \C$ then $(F^T, G^T)$ is an adjunction.

\ps{(cf.~\cite{linton:2}) 1. $G^T: \C^T \ar \C$ creates limits.
\\2. $G^T: \C^T \ar \C$ creates $G^T$-split coequalizers.
\\3. $F^T: \N \ar \C_T$ preserves any colimit in $\N$ which is also
a colimit in $\C$.
\\4. There is a bijection between morphisms $T
\ar T'$ of clones in extension form on $\N$ and functors $\C^{T'}
\ar \C^T$ of concrete categories over $\C$. }

\df{Suppose $T$ and $T'$ are two clones in extension form over two
full subcategories $\N$ and
 $\N'$ of $\C$ respectively.
\\1. $T$ and $T'$ are rational equivalent if $X^T$ and $X^{T'}$ are equivalent as
concrete category over $\C$.
\\2. $T$ and $T'$ are Morita equivalent if $X^T$ and $X^{T'}$ are equivalent as abstract
category. }

\lm{Suppose $\N \subset \N'$ and any object of $\N'$ is a retract of
an object of $\N$. Then any clone over $\N'$ in  extension form is
rationally equivalent to its restriction to $\N$. }

\lm{(i) Any clone $T$ in extension form over $\N$ determines a clone
over $\N$, called the clone of $T$.
\\ (ii) Any clone $T$ over a dense subcategory $\N$ defines a clone in extension form over $\N$ whose clone is $T$.}

\rk{Suppose $\C$ is a complete category. Let $A$ and $N$ be two
objects of $\C$. Let $A^{A^N}$ be the $A^N$-th power of $A$. For any
$\nu: N \ar A$ let $\diam f: A^{A^N} \ar A$ be the projection
determined by $\nu$. Consider the map $*-: (A^{A^N})^N \ar
(A^{A^N})^{(A^{A^N})}$ defined by $(*\alpha) (\diam \nu) = \diam
(\alpha (\diam \nu))$ for any $\alpha: N \ar A^{A^N}$ and $\nu \in
A^N$. Let $\eta: N \ar A^{A^N}$ be the map such that $\eta (\diam
\nu) = \nu$ for any $\nu: N \ar A$. Then $(A^{A^N}, \eta, *-)$ is a
clone over $N$ in extension form, called a \la{transformation
monad}. Note that $(A, \diam-)$ is an algebra of the clone
$(A^{A^N}, *-)$. Let $D$ be an object. Suppose $(A, \eta, *-)$ is a
clone over $N$ in extension form and $\theta: A \ar D^{D^N}$ is a
morphism. If $(D, (\theta \diam)-)$ is an $A$-algebra then $\theta:
(A, *-) \ar (D^{D^N}, *-)$ is a morphism of clones over $N$.
Conversely if $\theta:(A, *-) \ar (D^{D^N}, *-)$ is a morphism of
clones over $N$ then  $(D, (\theta \diam)-)$ is an $A$-algebra. It
follows that if $\C$ is complete then an $A$-algebra can also be
defined as an object $D$ together with a morphism of clones over $N$
from $A$ to $D^{D^N}$.}

\te{(Cayley's theorem for clones) Any clone $A$ in extension form
over an object $N$ of a complete category is a subclone of the
transformation clone $A^{A^N}$ over $N$. }

\section{Clones in Universal Algebra}\label{sec:ua}

In abstract universal algebra one studies clones over the set $\mN$
of positive integers. Since $\mN$ is dense in $\Set$, the notion of
clones and clones in extension form over $\mN$ are equivalent. Such
a clone can be defined in many different ways (see W. D.
Neumann~\cite{newmann:rep}, B. M. Schein, V. S.
Trohimenko~\cite{schtro:algmul}, B. Pareigis and H.
Rohrl~\cite{parho:llinear}, and Z. Luo~\cite{luo:1}~\cite{luo:2}).
The following two equivalent definitions stand out as the most
convenient definitions to use in practice:

If $A$ is any set denote by $A^{\mN}$ the set of infinite sequences
$[a_1, a_2, ...]$ of elements of $A$. If $A, B, C$ are three sets
and  $\alpha: A \times B^{\mN} \ar C$ is a function we often simply
write $a[b_1, b_2, ...]$ for $\alpha(a, [b_1, b_2, ...]) \in C$ for
any $a \in A$ and $b_1, b_2, ... \in B$. We shall use these
notations in the definitions of clones over $\mN$ and left or right
algebras of a clone over $\mN$.

\begin{defn}\label{def:clone2}
A clone over $\mN$ is a triple $(\A, X, \sigma)$ where $\A$ is a
nonempty set, $X = \{x_1, x_2, ...\}$ is a subset of $\A$, and
$\sigma: \A \times \A^{\mN} \ar \A$ is a map such that for any $a,
a_1, a_2, ..., b_1, b_2, ... \in \A$ we have
\\1. $(a[a_1, a_2, ...])[b_1, b_2, ...] = a[a_1[b_1. b_2, ...],
a_2[b_1, b_2, ...], ..]$.
\\ 2. $a[x_1, x_2, ...] = a$.
\\3. $x_i[a_1, a_2, ...] = a_i$ for any $i \in \mN$.
\end{defn}

\df{A clone over $\mN$ is a nonempty set $\A$ such that
\\1. $\A^{\mN}$ is a monoid with a unit $\tx = [x_1,
x_2, ...]$.
\\2. $\A$ is a right $\A^{\mN}$-act.
\\ 3. $x_i[a_1, a_2, ...] = a_i$ for any $\ta = [a_1, a_2, ...] \in
\A^{\mN}$ and $i \in \mN$.}

\rk{According to Definition \ref{def:clone2} a clone over $\mN$ is
an algebra with an $\mN$-ary operation $\alpha: \A^{\mN} \times \A =
\A^{\mN}  \ar \A$ and a countably infinite sequence of constants
$\{x_1, x_2, ...\}$ satisfying the three axioms. Thus the class of
clones over $\mN$ forms a (non-finitary) variety.}

From now on by a clone we always mean a \la{clone over $\mN$}. We
shall write $[[a_1, a_2, a_3, ..., a_n]]$ for $[a_1, a_2, a_3, ...,
a_n, a_n, a_n, ...]$ for any $a_1, a_2, ..., a_n \in \A$. Let $X =
\{x_1, x_2, ...\}$.

The notions of left and right algebras of clones on $\mN$ defined
below were first introduced in~\cite{parho:llinear}.

\df{Let $\A$ be a clone. A left algebra of $\A$ (or a left
$\A$-algebra) is a set $D$ together with a multiplication $\A \times
D^{\mN} \ar D$ such that for any $a \in \A$, $[a_1, a_2, ...] \in
\A^{\mN}$ and $[d_1, d_2, ...] \in D^{\mN}$ \\1. $(a[a_1, a_2,
...])[d_1, d_2, ...] = a([a_1[d_1, d_2, ...], a_2[d_1, d_2, ...],
...]$. \\2. $x_i[d_1, d_2, ...] = d_i$.}

\ex{1. $\mN$ is a clone with the monoid $\mN^{\mN}$.  It is the
initial clone.
\\2. Similarly $X =\{x_1, x_2, ...\}$ is an initial clone which is isomorphic to $\mN$.
\\3. Let $\tau = \{n_i\}_{i \in I}$ be a type of algebras. Let $X =
\{x_1, x_2, ...\}$ be a set of variables. The term algebra
$\T_{\tau}(X)$ defined in universal algebra is a clone. The category
of $\tau(X)$-algebra is equivalent to the category of left
$\T_{\tau}(X)$-algebras. }

\df{Let $\A$ be a clone. A right algebra of $\A$ (or a right
$\A$-algebra) is a right act $B$ of monoid $\A^{\mN}$. Suppose $\A$
and $\B$ are two clones. An $\A$-$\B$-algebra is a left $\A$-algebra
and right $\B$-algebra $B$ such that $(a\tc)\tb = a(\tc\tb)$ for any
$a \in \A$, $\tc \in B^{\mN}$ and $\tb \in \B^{\mN}$. }

\ex{$\A$ is an $\A-\A$-algebra.}

We say an element of $b$ of a right $\A$-algebra $B$  has \la{finite
rank} $n
> 0$ if $b[[x_1, x_2, ..., x_n]] = b$. We say $b$ has \la{rank
$0$} (or \la{$b$ is closed}, or \la{$b$ is a sentence}) if $b[[x_1]]
= b[[x_2]] = b$. An element $b \in B$ has rank $n \ge 0$ if and only
if the left translation $l_a: \A^{\mN} \ar B$ with $\ta \ar b\ta$
only depends on the first $n$ components of $\ta$. Denote by $\F(B)$
the set of finitary elements of $B$. Denote by $\F_n(B)$ the set of
elements of $B$ with rank $n \ge 0$. $B$ is \la{locally finitary} if
any element of $B$ has finite rank. We obtain a sequence of sets:
\[\F_0(B) \subseteq \F_1(B) \subseteq \F_2(B) \subseteq ... \F(B)
\subseteq B.\] Since $\A$ itself is also a right $\A$-algebra, these
notions also apply to $\A$. Hence we have a sequence of subsets of
$\A$:
\[\F_0(\A) \subseteq \F_1(\A) \subseteq \F_2(\A) \subseteq ... \F(\A)
\subseteq \A.\]

 \lm{1. $\F(\A)$ is a locally finitary
clone.
\\2. The category of locally finitary clone is a full coreflectvie subcategory
of the category of clones.
\\3. Each $\F_i(\A)$ is a free left $\A$-algebra of rank $i \ge 0$.
\\4. $\A$ is a free left $\A$-algebra of countable rank.
}

\df{The dull $Law(\A)$ of the full subcategory $\F_0(A) \subseteq
\F_1(A) \subseteq \F_2(A) \subseteq ... $ of the category of left
$\A$-algebras is called the Lawvere theory of $\A$. Note that
$Law(\A) = Law(\F(\A))$.}

\lm{Suppose $\A$ is a locally finitary clone.
\\1. If $B$ is a right $\A$-algebra then $\F(\B)$ is a locally finitary right
$\A$-algebra.
\\2. The category of locally finitary right $\A$-algebras is a full
coreflectvie subcategory of right $\A$-algebras. }

\te{1. If $\A$ is a locally finitary clone then the class of left
$\A$-algebras is a finitary variety. Conversely any finitary variety
arises in this way.
\\2. The category of locally finitary clone is equivalent to
the opposite of the category of finitary varieties (as concrete categories over $\Set$).
\\3. The category of locally finitary clone is equivalent to the category of
Lawvere algebraic theories (without terminate object).
\\4. The category of locally finitary clone is equivalent to the category of
finitary monads in $\Set$. }

\df{Let $V$ be a variety of type $\tau$. Let $\A$ be a clone. An
$(\A, V)$-algebra is a set $B$ which is  a right $\A$-algebra and an
algebra in $V$ such that for any $n$-ary operation symbol $f$ in
$\tau$ the operation $f^B: B^n \ar B$ is a homomorphism of right
$\A$-algebras, or equivalently, each action on $B$ induces an
endomorphism on $B$. If $B = \A$ then we say $\A$ is a $V$-clone. If
$V$ is the variety of all $\tau$-algebras then an $(\A, V)$-algebra
or $V$-clone is called an $(\A, \tau)$-algebra or $\tau$-clone
respectively. Denote by $Clone$ the variety of clones. A
$Clone$-clone is called a $C$-clone.}

\df{A $V$-clone (or $\tau$-clone) $\A$ is called primary if the
algebra $\A$ in $V$ is generated by $X$.}

\lm{1. If $\A$ is a primary $\tau$-clone then the class of left
$\A$-algebras is a $\tau$-variety with $\A$ as the free algebra over
$X$.

2. The category of primary $\tau$-clones is equivalent to the
opposite of the category of $\tau$-varieties. }

\ex{1. Suppose $T$ and $D$ are two left algebras of clones $\A$ and
$\B$ respectively. Then the set of maps $T^{D^{\mN}}$ from $D^{\mN}$
to $T$ is an $\A-\B$-algebra.
\\2. Suppose $D$ is a left $\A$-algebra and $T$ is an algebra in a
variety $V$, then  $T^{D^{\mN}}$ is an $(\A, V)$-algebra. }

\ex{Suppose $B$ is an algebra in a variety $V$. Then the basic
operations on $B$ extend to $B^{B^{\mN}}$ point-wisely, so
$B^{B^{\mN}}$ is a $V$-clone. Denote by $Cl(B)$ the subalgebra of
$B^{B^{\mN}}$ generated by the projections $\pi_1, \pi_2, ...$ from
$B^{\mN}$ to $B$. One can show by induction that $Cl(B)$ is a
subclone of $B^{B^{\mN}}$, thus it is the smallest sub-$V$-clone of
$B^{B^{\mN}}$.  If $V$ is a finitary variety then $Cl(B)$ is a
locally finitary clone, and the Lawvere theory $Law(Cl(B))$ of
$Cl(B)$  determines the clone of $B$ in the classical sense of P.
Hall (cf.~\cite{cohn:1} p.126). If $B$ is a clone then $Cl(B)
\subseteq B^{B^{\mN}}$ are $C$-clones.}

A $C$-clone can also be defined directly:

\df{A \la{$C$-clone} is an algebra $\Hy$ with two $\mN$-ary
operations $^., \ *: \Hy \times \Hy^{\mN} = \Hy^{\mN} \ar \Hy$ and
two sequences $t_1, t_2, ... , s_1, s_2, .... \in \Hy$ such that
\\1. $(\Hy, \ ^., \{t_1, t_2, ...\})$ and $(\Hy, *, \{s_1, s_2,
...\})$ are clones.
\\2. $(au)*v = (a*v)[u_1*v, u_2*v, ...]$ for any $a \in \Hy$ and $u,v
\in \Hy^{\mN}$.
\\3. $t_i*u = t_i$ for any $u
\in \Hy^{\mN}$.}

The class of $C$-clones is a variety. Therefore the initial
$C$-clone $E$ exists. It is the free clone on $\{s_1, s_2, ...\}$.
For any $C$-clone $\Hy$ the unique homomorphism $L_{\Hy}: E \ar \Hy$
is not injective in general, which determines a congruence
$Cng(\Hy)$ of $E$. We say that \la{a $C$-clone $\Hy$ satisfies a
hyperidentity $a \approx b$} if $<a, b> \ \in Cng(\Hy)$. We say that
\la{a clone $\A$ satisfies a hyperidentity $a \approx b$} if the
$C$-clone $B^{B^{\mN}}$ (or $Cl(\A)$) satisfies the hyperidentity.
We say that \la{a variety $V$ satisfies a hyperidentity $a \approx
b$} if the clone determined by the free algebra of $V$ on $\{t_1,
t_2, ...\}$ satisfies the hyperidentity. Thus we may speak of the
hyperidentities satisfied by groups, rings, etc. The theory of
hyperidentities are important for the study of classification of
finitary varieties. A \la{hypervariety} is the totality of locally
finitary clones satisfying a set of hyperidentities. A class of
locally finitary clones is a hypervariety if and only if it is
closed under subclones, homomorphic images and direct products of
clones (cf.~\cite{de:1}).

\ex{ (Post) Let $n$ be any positive integer $> 1$ viewed as a finite
set with $n$ elements. Then $\F(n^{n^{\mN}})$ is a locally finitary
clone. The category of left $\F(n^{n^{\mN}})$-algebras is equivalent
to the category of boolean algebras. In particular, the category of
left $\F(2^{2^{\mN}})$-algebras is equivalent to the category of
boolean algebras. Note that $\F(2^{2^{\mN}})$ is naturally a free
boolean algebra of countable rank.}

\ex{ Let $\mF = \{0, 1, 2, 3, ...\}$  and  $\mF' = \{1, 2, 3, ...\}$
be the full subcategories of $\Set$. Then the category of locally
finitary right $X$-algebras is equivalent to the category of
presheaves in $\Set^{\mF'}$. Since each presheaf in $\Set^{\mF'}$ is
the normalization of a presheaf in $\Set^{\mF}$ at $0$, one may
replace presheaves in $\Set^{\mF}$ by locally finitary right
$X$-algebras (or more generally, by any right algebra of a clone) in
the definition of binding algebras over variable sets given
in~\cite{pl:1}.}

\df{A type is a right $X$-algebra.}

Denote by $Type$ (resp. $FType$) the category of types (resp.
locally finitary types). Then $FType$ is a full coreflective
subcategory of $Type$.

Suppose $D$ and $E$ are two locally finitary types. Then $E^{\mN}$
is a bi-act of $\mN^{\mN}$. The tensor product $D \otimes E^{\mN}$
is the quotient of the product type $D \times E^{\mN}$ modulo the
congruence generated by the relations $(d[y_1, y_2, ...], [e_1, e_2,
...]) \approx (d, [y_1, y_2, ...][e_1, e_2, ...])$ for all $d \in
D$, $e_1, e_2, ... \in E$ and $y_1, y_2, ... \in X$. Then $D \otimes
E^{\mN}$ is a locally finitary type. One can show that $(FType,
\otimes, X)$ is a strict monoidal category with the unit $\mN$.

Any locally finitary type $D$ determines a finitary endofunctor
$\eta_D: \Set \ar \Set: E \ar D \otimes E$ (here each set $E$ is
viewed as a type with trivial actions). Conversely, if $F: \Set \ar
\Set$ is any finitary endofunctor then $F(\mN)$ is naturally a
locally finitary type.

\te{1. The category of strict monoidal category of locally finitary
types is equivalent to the strict monoidal category of finitary
endofunctors of $\Set$ with compositions as multiplications.

2. The category of locally finitary clones is equivalent to the
category of monoids in the monoidal category $(FType, \otimes, X)$.
}

\ex{If $\A$ is a clone and $Y$ is any set then $\A \otimes Y$ is
naturally a left $\A$-algebra which is the free left $\A$-algebra
over $Y$.}

\te{ If $\A$ is any clone denote by $\E(\A^{\mN})$ the set $\tee$ of
idempotents of $\A^{\mN}$ such that $\tee = [[x_1, x_2, ...,
x_n]]\tee[[x_1, x_2, ..., x_n]]$ for some $n > 0$,  which is viewed
as a subcategory of Karoubi envelope of $\A^{\mN}$ (cf.~\cite{ba:1}
p.114). Then two locally finitary clones $\A$ and $\B$ are Morita
equivalent iff the categories $\E(\A)$ and $\E(\B)$ are equivalent
(cf.~\cite{ad:1}) (note that bi-algebras of clones can also be used
to characterize Morita equivalent clones) .}

Suppose $\A$ is a clone. Let $n > 0$ be a positive integer. If $u =
(a_1, ..., a_n), v = (b_1, ..., b_n) \in \A^n$ we write $u + v$ for
the join $(a_1, ..., a_n, b_1, ..., b_n)$. The \la{ $n$-th metrix
power of $\A$} is the clone $\A^{[n]}$ with $\A^n$ as the universe
such that the new multiplication $\A^n \times (\A^n)^{\mN} \ar \A^n$
is defined by $u[v_1, v_2, ...](i) = u(i)[v_1 + v_2 + ...]$ for all
$u, v_1, v_2, ... \in \A^n$ and $i \leq n$. The unit of $\A^{[n]}$
is \[[(x_1, ..., x_n), (x_{n+1}, ..., x_{2n}), (x_{2n +1},
...,x_{3n}), ... )].\] Using the above theorem one can show that
$\A$ and $\A^n$ are Morita equivalent for locally finitary $\A$.

There is yet another standard method to obtain Morita equivalent
clones, called \la{modification}. Call an element $a \in \A$ a
\la{varietal generator (of rank 1)} if $a = a[[x_1]] = a[[a]]$ and
$[[x_1]] = \tc([[a]][[x_n]])\td$ for some $\ta, \td \in \A^{\mN}$
and $n > 0$. If $a$ is a varietal generator
 then $a\A^{\mN}$ is a clone with $[a[[x_1]], a[[x_2]],
...]$ as the unit. Using the above theorem again one can show that
$\A$ and $a\A^{\mN}$ are Morita equivalent.

\te{A locally finitary clone $\B$ is Morita equivalent to a locally
finitary clone $\A$ iff $\B \cong a(\A^{[n]})^{\mN}$ for a varietal
generator $a$ of $\A^{[n]}$ for some $n > 0$ (cf.~\cite{mc:1}).}

There are many other interesting aspects of clone theory. Since
Lawvere theories, finitary varieties and finitary monads over $\Set$
are essentially locally finitary clones over $\mN$, any notion
applies to Lawvere theories, finitary varieties and finitary monads
over $\Set$ can be transformed into a purely algebraic notion for
locally finitary clones (or their left or right algebras).
Furthermore, if a notion does not specifically refer to the
finiteness condition then it can also be extended to arbitrary
clones. Here are some examples: tensor product of clones,
commutative clone, Morita theory for clones, Malcev clone,
arithmetical clone, minimal clone, discriminator clone, spectrum of
a clone, Post algebra, cylindric algebra or polyadic algebra with
terms in a clone, Fiore-Plotkin-Turi substitution algebra, and
operads, etc.

\section{Algebraic Theories}\label{sec:at}

\df{1. An algebraic theory is a category $(T, T^{\mN})$ of two
objects such that $T^{\mN}$ together with a map $x: \mN \ar
\hom(T^{\mN}, T)$ is the $\mN$-th power of $T$.
\\2. Suppose $(T, T^{\mN})$ and $(H, H^{\mN})$ are two algebraic
theories. A functor $F: (T, T^{\mN}) \ar (H, H^{\mN})$ is called a
morphism of algebraic theories if $F$ sends $T$ to $H$, $T^{\mN}$ to
$H^{\mN}$ and $F(x(i)) = x(i)$ for any $i \in \mN$.
\\3. A left model of an algebraic theory $(T, T^{\mN})$ is a functor $(T, T^{\mN})
\ar \Set$ preserving the $\mN$-th power of $T$. A homomorphism of
left models of $(T, T^{\mN})$ is a natural transformation.
\\4. A right model of $(T, T^{\mN})$  is a functor $(T, T^{\mN})
^{op} \ar \Set$. A homomorphism of right models of $(T, T^{\mN})$ is
a natural transformation. }

\rk{1. Any clone $\A$ determines an algebraic theory $(\A,
\A^{\mN})$, which is the full subcategory of right acts of
$\A^{\mN}$ generated by the two right acts $\A$ and $\A^{\mN}$. The
algebraic theory $(\A, \A^{\mN})$ is called a \la{matrix algebraic
theory}.
\\2. Any algebraic theory $(T, T^{\mN})$ determines a clone
$\hom(T^{\mN}, T)$  with the monoid $\hom(T^{\mN}, T)^{\mN} =
\hom(T^{\mN}, T^{\mN})$ and the unit $x: \mN \ar \hom(T^{\mN}, T)$.
\\3. These processes are inverse to each other. Thus the notion of
clones (over $\mN$) is equivalent to the notion of algebraic
theories. More precisely, we have the following}

\lm{1. Any algebraic theory $(T, T^{\mN})$ is isomorphic to the
matrix algebraic theory $(\hom(T^{\mN}, T), \hom(T^{\mN},
T)^{\mN})$.
\\2. The category of clones (over $\mN$) is equivalent to the category of algebraic
theories. }

\ex{Let $V$ be a variety. Let $F(1)$ and $F(\mN)$ be the free
algebras of rank $1$ and $\mN$ respectively. Then $F(\mN)$ is the
$\mN$-th sum of $F(1)$. Thus the dual of the subcategory $(F(1),
F(\mN))$ of $V$ is an algebraic theory, and $F(\mN) = hom(F(1),
F(\mN))$ is a clone, called the \la{clone of $V$}. For instance, if
$S$ is any set then $(S, \mN \times S)^{op}$ is an algebraic theory.
Thus $(\mN \times S)^S$ is a clone. }

\section{Binding Operations}\label{sec-bo}
Suppose $\A$ is a clone and $B$ is a right $\A$-algebra. Denote by
$\tx = [x_1, x_2, ...]$ the unit of the monoid $\A^{\mN}$.

If $a \in \A$ and $\tb \in B^{\mN}$ we let
\\$[+] = [x_2, x_3, x_4, x_5, ...] \in \A^{\mN}$.
\\$[-] = [x_1, x_1, x_2, x_3, x_4, x_5, ...] \in \A^{\mN}$.
\\$[+i] = [x_1, x_2, ..., x_{i-1}, x_{i+1}, ...] \in \A^{\mN}$.
\\$[-i] = [x_1, x_2, ..., x_{i-1}, x_i, x_i, x_{i+1}, ...] \in \A^{\mN}$.
\\$a^+ = a[+]$, $a^- = a[-]$, $a^{+i} = a[+i]$, $a^{-i} = a[-i]$, $\tb^{+i} =
\tb[+i]$, $\tb^{-i} = \tb[-i]$.

In the following we assume $y, z, w, ... \in \{x_1, x_2, ...\}$,
which are called \la{syntactical variables}. If $a, b \in B$ let
$a[b/x_i] = a[x_1, x_2, ..., x_{i-1}, b, x_{i+1}, ...]$. We say $a
\in B$ is \la{independent of a syntactical variable $y$} if $a =
a[y^+/y]$; otherwise we say that $a$ \la{depends on $y$}. Denote by
$FV(a)$ the set of variables on which $a$ depends. If $a$ has rank
$n \ge 0$ then $FV(a) \subseteq \{x_1, x_2, ..., x_n\}$. If $a$ is
closed then $FV(a) = \emptyset$. If $a$ is finitary and $FV(a) =
\emptyset$ then $a$ is closed.

\df{A map $\lambda: B \ar B$ is called an \la{abstract binding
operation on $x_i$} if \[(\lambda b)\ta = \lambda (b[a_1^{+i},
a_2^{+i}, ..., a_{i-1}^{+i}, x_i, a_i^{+i}, ...])\] for any $b \in
B$ and $\ta \in A^{\mN}$.}

\df{An operation $\lambda: B \ar B$ is called  \la{binding on $x_i$}
if
\[(\lambda b)\ta = (\lambda (b[a_1^{+i}, a_2^{+i}, ...,
a_{i-1}^{+i}, x_i, a_{i+1}^{+i}, ...]))^{[-i]}\] for any $b \in B$
and $\ta \in A^{\mN}$. }

\lm{1. If $\lambda: B \ar B$ is an abstract binding operation on
$x_i$ then the operation $\lambda^{+i}: B \ar B$ sending $b$ to
$(\lambda b)^{+i}$ is binding on $x_i$.
\\2. If $\lambda: B \ar B$ is binding on $x_i$ then the map $\lambda^{-i}: B \ar B$ sending
$b$ to $(\lambda b)^{-i}$ is an abstract binding operation on $x_i$.
\\3. The set of abstract binding operations on $x_i$ and
 the set of binding operations on $x_i$ for $B$ are bijective. }

\rk{1. A map $\lambda: B \ar B$ is an abstract binding operation on
$x_1$ if and only if $(\lambda b)\ta = (\lambda (b[x_1, a_1^+,
a_2^+, ...]))$.
\\2. A map $\lambda: B \ar B$ is binding on $x_1$ if and only if
$(\lambda b)\ta = (\lambda (b[x_1, a_2^+, a_3^+, ...]))^-$. }

\df{Let $\lambda: B \ar B$ be an abstract binding operation on
$x_1$. For any variable $y = x_i$ we introduce a new map $\ld y: B
\ar B$ by
\begin{align*}
&\ld y.b = \lambda (b^+[x_1/y^+]) = \lambda (b[x_2, x_3, ...,y^-, y,
x_1, y^{++}, ....]) \\& = \ld x_i.b = \lambda (b^+[x_1/x_i^+]) =
\lambda (b[x_2, x_3, ..., x_i, x_1, x_{i+2}, ....]).
\end{align*}}

\lm{If $j \le i$ then $\ld x_i.b =  (\ld
x_j.(b^{+j}[x_j/x_i^+]))^{-j}$.
\\ If $i < j$ then $\ld x_i.b = (\ld
x_j.(b^{+j}[x_j/x_i]))^{-j}$.}

Suppose $\lambda: B \ar B$ is an abstract binding operation on $x_1$
and $b \in B$.

\lm{1. $\ld x_1.b = \lambda(b[x_1, x_3, x_4, ...]) = (\lambda b)^+$.
\\2. $\lambda b = (\ld x_1.b)^-$.
 \\3. $(\ld x_1. b)\ta = (\ld x_1.(b[x_1, a_2^+,
a_3^+, ...]))^-$
\\4. $\ld x_1. b = (\ld x_1.(b[x_1, x_3, x_4,
...]))^-$.
\\5. $\ld x_i.b =  (\ld x_1.(b[x_2, x_3, ..., x_{i-1}, x_i, x_1, x_{i+2}, ..]))^-$.
\\6. $\ld y.b =  (\ld x_1.(b[x_2, x_3, ..., y^-, y, x_1, y^{++}, ..]))^-$.
\\7. $ (\ld x_i.b)[u_1, u_2, ...] = \lambda(b[u_1^+, u_2^+, ...,
u^+_{i-1}, x_1, u^+_{i+1}, u^+_{i+2}, ...])$.
\\8. $\ld y$ is binding on $y$.
\\9. $\lambda (\F_n(B)) \subset \F_{n-1}(B)$ for any
$n > 0$.
 \\ 10. $\lambda (\F_0(B)) \subset \F_0(B)$.
\\11. If $a \in B$ has finite rank $i > 0$ then $\lambda^i a$ is closed.
}

\lm{1. If $b$ is independent of $y$ then $ \ld z.b = \ld y.
(b[y/z])$.
\\2. If $b$ is independent of $y$ then  $\ld y.b = \lambda (b^+)$.
\\3. If $y \ne z$ and $a \in \A$ is independent of $z$ then $(\ld z.b)[a/y] =
\ld z.(b[a/y])$.
\\4. If $b$ is independent of $x_2$ then $(\ld x_1. b)^- = \ld x_1.
(b^-)$.
\\5. If $a_2, a_3, ...$ are independent of $x_1$ then $(\ld
x_1. b)\ta = (\ld x_1.(b[x_1, a_2^+, a_3^+, ...]))^- = \ld
x_1(b[x_1, a_2, a_3, ...])$. }

\rk{We have \\$(\ld x_1. x_1)^-  = \ld x_1.x_1 = \lambda x_1$
(closed),
\\$(\ld x_1.x_2)^-  =  \ld x_i.x_1 = \lambda x_2$ (rank 1) for $i > 2$.
\\$ (\ld x_1. x_3)^-  = \ld x_1.x_2 = \lambda x_3$ (rank 2),
\\$ (\ld x_1. x_4)^-  =  \ld x_1.x_3 = \lambda x_4$ (rank 3),
\\ ...
\\The irregularity for $(\ld x_1. x_2)^-$ is due to the fact that the substitution
$[x_1, x_1, x_3, ...]$ replaces $x_2$ by $x_1$, while in $\ld
x_1.x_2$ the variable $x_2$ is bound by $x_1$.}

\section{Clones in Lambda Calculus. }\label{sec-lc}
Suppose $\A$ is a clone. Let $B$ be a right $\A$-algebra, i.e. $B$
is a right act of the monoid $\A^{\mN}$.  We define a new right
$\A$-algebra $B^{\A} = (B, *)$ with the new action $*: B \times
\A^{\mN} \ar B$: $b*[a_1, a_2, ....] = b[x_1, a_1^+, a_2^+, ...]$. A
map $\lambda: B^{\A} \ar B$ is a homomorphism of right $\A$-algebras
if and only if $(\lambda b)\ta = \lambda(b * \ta) = \lambda(b[x_1,
a_1^+, a_2^+, ...])$ for any $\ta \in \A^{\mN}$. Thus $\lambda: B
\ar B$ is a homomorphism $B^{\A} \ar B$ if and only if it is an
abstract binding operation on $x_1$. Let $ev_{\A, B}: B^{\A} \times
\A \ar B$ be the homomorphism of right $\A$-algebras defined by
$ev(b, a) = b[a, x_1, x_2, ...]$ for any $b \in B$ and $a \in \A$.
The right $\A$-algebra $B^{\A}$ together with the homomorphism
$ev_{\A, B}: B^{\A} \times \A \ar B$ is the exponent in the category
of right $\A$-algebras. Specifically, this means that, for any $f: T
\times \A \ar B$ of homomorphism of right $\A$-algebras, there is a
unique $\Lambda f: T \ar B^{\A}$ (called the \la{curred version of
$f$}) given by $(\Lambda f)t = f(t^+, x_1)$ such that $f = ev \circ
(\Lambda f \times id_{\A})$, as we have $ev \circ (\Lambda f \times
id_{\A})(t, a) = ev_{\A, B}(f(t^+, x_1), a) = f(t^+, x_1)[a, x_1,
x_2, ...]  = f(t^+[a, x_1, x_2, ..], x_1[a, x_1, x_2, ...]) = f(t,
a)$.

The most important property about a clone is that it is a
\la{Kleisli algebra}, i.e. a monad over a category with only one
object. Algebraically a Kleisli algebra is a set $S$ together with
two monoid structures $(S, \ ^.)$ and $(S, \circ)$ such that $a(b
\circ c) = (ab) \circ c$ for any $a, b, c \in S$ (cf~\cite{mane} p.
110. ex.18 and p.136. ex 5). Suppose $\A$ is a clone. $\A^{\mN}$
carries another monoid structure with the binary operation $\circ$
in $\A^{\mN}$ defined by \[\ta \circ \tb = \ta[x_1, b_1, b_2, ...] =
[a_1[x_1, b_1, b_2, ...], a_2[x_1, b_1, b_2, ...], ...],\] whose
unit is $[+] = [x_2, x_3, ...]$. Denote by $(\A^{\mN}, \circ)$ this
new monoid. There are three basic homomorphisms of
monoids:\begin{center} $\Delta_1 : \A^{\mN} \ar (\A^{\mN}, \circ) \
\  [a_1, a_2, ...] \ar [a_1^+, a_2^+, ...]$,\\$\Delta_2: (\A^{\mN},
\circ) \ar \A^{\mN} \ \ [a_1, a_2, ...] \ar [x_1, a_1, a_2, ...]$.\\
$\Delta = \Delta_2 \Delta_1 : \A^{\mN} \ar \A^{\mN}  \ \ [a_1, a_2,
...] \ar [x_1, a_1^+, a_2^+, ...]$.\end{center} Then $\Delta_1$ is a
left adjoint of $\Delta_2$, which induces a monad $(\Delta, [+],
[x_1, x_1, x_2, ...]) $ on the one object category $\A^{\mN}$.  We
have $\ta(\tb \circ \tc) = (\ta\tb) \circ \tc$ for $\ta, \tb, \tc
\in \A^{\mN}$. Thus $(\A, \ ^., \ \circ)$ is a Kleisli algebra.
Denote by $Rg(\A)$ the category of right $\A$-algebras. Let $\delta:
Rg(\A) \ar Rg(\A)$ be the functor induced by the homomorphism
$\Delta: \A^{\mN} \ar \A^{\mN}$. Then for any right $\A$-algebra $B$
we have $\delta(B) = B^{\A}$.

\ex{The submonoid of $\mN^{\mN}$ generated by $[+]$ and $[-]$ is the
initial Kleisli Algebra.}

Let $\C$ be a cartesian closed category. An object $U \in \C$ is
\la{reflexive} if the exponent $U^U$ is a retract of $U$, i.e. there
are maps $F: U \ar U^U$ and $G: U^U \ar U$ such that $FG = id_{U^U}
$.

\df{1. A \la{reflexive clone} is a clone $\A$ together with two
homomorphisms $\lambda: \A^{\A} \ar \A$ and $\lambda^*: \A \ar
\A^{\A}$ such that $\lambda^* \lambda  = id_{\A^{\A}}$.
\\2. An \la{extensional clone} is a clone $\A$ together with a
bijective homomorphism (i.e. an isomorphism) from $\A$ to $\A^{\A}$
(cf~\cite{ba:1}). }

\df{(Elementary Definition) A \la{reflexive clone} (resp.
\la{extensional clone}) is a clone $\A$ together with two maps
$\lambda, \lambda^*: \A \ar \A$ such that
\\1. $\lambda^* (a\ta) = (\lambda^* a)[x_1, a_1^+, a_2^+, ...]$.
\\2. $(\lambda a)\ta = \lambda(a[x_1, a_1^+, a_2^+, ...])$ (resp.
$\lambda \lambda^*  = id_{\A}$).
\\3. $\lambda^* \lambda  = id_{\A}$.
}

\lm{1. If $\A$ is a reflexive clone then $\F(\A)$ is a locally
finitary reflexive clone.
\\2. The category of locally finitary reflexive clones is a coreflective
subcategory of the category of reflexive clones.
\\3. The initial reflexive clone is locally finitary.
The same is true for extensional clones.}

\te{ Any reflexive object $U$ in a cartesian closed category
determines a reflexive clone.} \pf{1. First assume the $\mN$-th
power $U^{\mN}$ of $U$ exists. Let $\A = \hom(U^{\mN}, U)$.  Since
$U^U$ is a retract of $U$ and $U$ is a retract of $U^{\mN}$,
$U^{\mN}$ is dense in the subcategory $(U^U, U, U^{\mN})$ of $\C$.
Thus $(U^U, U, U^{\mN})$ is equivalent to the category $(\A^{\A},
\A, \A^{\mN})$ of right acts of $\A^{\mN}$. Thus $\A$ is a reflexive
right act.
\\2. If $U^{\mN}$ dose not exist then one can embed the Lawvere
theory $(U^0, U, U^2, U^3, ...)$ in the opposite of the category of
its models. Then $U$ is reflexive and the $\mN$-th power $U^{\mN}$
of $U$ exists. Applying step 1 we obtain a reflexive clone. }

Suppose $\lambda^*: \A \ar \A^{\A}$ is a homomorphism. We have
$\lambda^* a_1 = \lambda^* (x_1\ta) = (\lambda^* x_1)[x_1, a_1^+,
a_2^+, ...] = (\lambda^* x_1)[[x_1, a_1^+]]$. Thus $\lambda^* a =
(\lambda^* x_1)[[x_1, a^+]]$ and $\lambda^*(x_1) \in \F_2(\A)$.
Define a homomorphism $\A \times \A \ar \A$ of right $\A$-algebras
by $ab = (\lambda^* x_1)[[b, a]]$. Then $\lambda^* a = a^+x_1$.
Conversely if $f: \A \times \A \ar \A$ is a homomorphism of right
$\A$-modules then the map $\lambda^*: \A \ar \A$ defined by $a \ar
a^+x_1$ is a homomorphism from $\A$ to $\A^{\A}$.

\lm{The following three sets are bijective:
\\1. $\hom(\A, \A^{\A})$.
\\2. The set of homomorphisms $\A \times \A \ar \A$ of right $\A$-algebras.
\\3. $\F_2(\A)$. }

\df{A $\lambda$-clone is a clone with the following homomorphism of right $\A$-algebras: \\
(i) $\lambda: \A^{\A}  \ar \A$.
\\(ii) An application $^.: \A \times \A \ar \A$ (or equivalently, a homomorphism $\lambda^*: \A \ar
\A^{\A}$).
\\A  $\lambda_{\beta}$-clone is a  $\lambda$-clone satisfying the
following axiom for any $a \in \A$ \\(a) $(\lambda a)^+x_1 = a$ (or
equivalently, $\lambda^*\lambda = id_{\A}$) ($\beta$-conversion).
\\A  $\lambda_{\beta \eta}$-clone is a  $\lambda_{\beta}$-clone
satisfying the following axiom for any $a \in \A$ \\ (b) $\lambda
(a^+x_1) = a$ (or equivalently, $\lambda\lambda^* = id_{\A}$)
($\eta$-conversion).}

The category of reflexive clones (resp. extensional clones) is
equivalent to the category of $\lambda_{\beta}$-clones (resp.
$\lambda_{\beta \eta}$-clones). The classes of $\lambda$-clones,
$\lambda_{\beta}$-clones, and $\lambda_{\beta \eta}$-clones are
varieties in the sense of universal algebra. Traditionally the
initial $\lambda$-clone
     $\Lambda$ is defined by
$\lambda$-terms:

\begin{defn}
 ($\lambda$-terms) The class $\Lambda$ of $\lambda$-terms is
the least class satisfying the following \nl 1. $x_i \in \Lambda$
for any $i \in \mN$. \nl 2. if $a \in \Lambda$ then $(\lambda a) \in
\Lambda$ \nl 3. if $a, b \in \Lambda$ then $(ab) \in \Lambda$
\end{defn}

We define a multiplication $\Lambda \times \Lambda^{\mN} \ar
\Lambda$ inductively: \nl 1. $x_i[a_1, a_2, ...] = a_i$. \nl 2.
$(ab)[a_1, a_2, ...] = (a[a_1, a_2, ...])(b[a_1, a_2, ...])$. \nl 3.
$(\lambda a)[a_1, a_2, ...] = \lambda(a[x_1, a_1^+, a_2^+, ...])$.
\\ One can prove by induction that $a[x_1, x_2, ...] = a$, and the
following lemma

\begin{lem} {\bf Substitution  Lemma.} $ (a\ta)\tb = a[a_1\tb, a_2\tb,
...]$.\end{lem}

It is easy to see that $\Lambda$ is a $\lambda$-clone. Clearly it is
the initial $\lambda$-clone.

\begin{defn}
1. Let $\Lambda_{\beta}$ be the quotient of the $\lambda$-clone
$\Lambda$ by the congruence generated by all the pairs $\{<a,
(\lambda a)^+x_1>\}_{a \in \Lambda}$. $\Lambda_{\beta}$ is the
initial $\lambda_{\beta}$-clone.
\\2. Let $\Lambda_{\beta \eta}$ be the quotient of the $\lambda$-clone
$\Lambda$ by the congruence generated by all the pairs $\{<a,
(\lambda a)^+x_1>\}_{a \in \Lambda}$ and $\{<a,
\lambda(a^+x_1)>\}_{a \in \Lambda}$. Then $\Lambda_{\beta \eta}$ is
the initial $\lambda_{\beta, \eta}$-clone. \\(Note that it follows
from Church-Rosser theorem that $\Lambda_{\beta}$ and
$\Lambda_{\beta \eta}$ are not trivial (i.e. they contains more than
one elements)~\cite{ba:1}).
\end{defn}

\df{1. A left $\Lambda$-algebra is simply called a
$\Lambda$-algebra.
\\2. A left $\Lambda_{\beta}$-algebra is called a
$\lambda$-algebra.
\\3. A left $\Lambda_{\beta \eta}$-algebra is called an extensional
$\lambda$-algebra.
\\4. A left $\Lambda_{\beta}$-algebra $D$ is called a $\lambda$-model if the following axiom is satisfied:
\\ if $a, b \in \A$ and $a\tm = b\tm$ for every $\tm \in D^{\mN}$ then $\lambda a = \lambda b$. }

Since $\Lambda$ is finitary with $\Lambda_{\beta}$ and
$\Lambda_{\beta \eta}$ as quotients, these two classes are also
finitary. Hence the classes of $\Lambda$-algebras (resp.
$\lambda$-algebras, resp. extensional $\lambda$-algebras) are
finitary varieties. Note that in a $\lambda$-clone we have the
derived unary operations $(\ld x_1), (\ld x_2), (\ld x_3), ...$
Thus the classical $\lambda$-terms can be interpreted in any
$\lambda$-clone. In fact, the initial $\lambda$-clone $\Lambda$ is
precisely the quotient of classical $\lambda$-terms modulo
$\alpha$-conversion.

\lm{ Suppose $a, b \in  \Lambda$.
\\1. $FV(x) = \{x\}$.
\\2. $FV(\ld x.a) = FV(a) - \{x\}$.
\\3. $FV(ab) = FV(a) \cup FV(b)$. }

\begin{exm} Suppose $\A$ is a $\lambda$-clone. \\ 1. $\ld y.y = \lambda ((y^+)[x_1/y^+]) = \lambda x_1$.
\\ 2.  $\ld y.y = \ld z.z$ for any variables $y$ and $z$.
\\3. If $\A$ is a $\lambda_{\beta}$-clone then $(\ld y.y)b = b$
for any $b \in A$.
\\4. If $i > 0$ then $\lambda x_i$ has a rank $i-1$, and $\lambda^i
x_i$ is closed.
\end{exm}

\rk{If $\A$ is a $\lambda_{\beta}$-clone then $(\ld y.a)b = a[b/y]$.
By induction one can see that this rule is sufficient for the
calculations in $\Lambda(\A)$ using the derived binding operations
$\{\ld y\}$ and simple substitutions $\{[b/y]\}$.}

A $\lambda$-clone can also be defined directly without referring to
the unary map $\lambda$:

\df{(Alternate Definition) A $\lambda$-clone is a clone with the following maps: \\
(i) $\ld x_1: \A  \ar \A$ such that $(\ld x_1. a)[a_1, a_2, ...] =
(\ld x_1.(a[x_1, a_2^+, a_3^+, ...]))^-$.
\\(ii) $^.: \A \times \A \ar \A$ is a homomorphism of right $\A$-algebras (called the application).
\\A  $\lambda_{\beta}$-clone is a  $\lambda$-clone satisfying the
following axiom for any $a \in A$ \\(a) $(\ld x_1. a)x_1 = a$
($\beta$-conversion).
\\A  $\lambda_{\beta \eta}$-clone is a  $\lambda_{\beta}$-clone
satisfying the following axiom for any $a \in A$ \\ (b) $\ld x_1.
(a^+x_1) = a^+$ ($\eta$-conversion).}

\section{Clones in Predicate Logic}\label{sec:fo}
In this section we present an ad hoc approach to first-order logic
to show how to eliminate the problems related to the complicated
notion of variable substitutions encountered in most traditional
approaches. We choose E. Mendelson~\cite{me:1} as our main
reference. All definitions are given in traditional fashion.
Although a few properties are stated in terms of clones, it is
straightforward to replace them by checking the properties directly.
Historically A. Tarski~\cite{ta:2} seems to be the first to address
these problems.

Let $\tau_p = \{\FF, \Rightarrow\}$ be a type of algebras, where
$\FF$ is a $0$-ary operation and $\Rightarrow$ is a binary
operation. Any $\tau_p$-algebra is called a \la{proposition
algebra}. For instance, $2 = \{0, 1\}$ is a proposition algebra with
$\FF = 0$ and $(0 \Rightarrow 0) = (0 \Rightarrow 1) = (1
\Rightarrow 1) = 1$, $(1 \Rightarrow 0)  = 0$. If $P$ is a
proposition algebra we introduce some further operations: $\neg p =
(p \Rightarrow \FF)$, $\TT = \neg \FF$, $p \vee q = (\neg p)
\Rightarrow q$, $p \wedge q = \neg (p \Rightarrow \neg q)$, $p
\Leftrightarrow q = (p \Rightarrow q) \wedge (q \Rightarrow p)$.

\df{A first-order language $\mL$ consisting of \\
\ (i) The set of individual variables $\{x_1, x_2, ...\}$. \\ \ (ii)
The set $\{\bF_n\}_{n \in \mN}$ of function symbols. \\ \ (iii) The
set $\{\bR_n\}_{n \in \mN}$ of predicate symbols. We assume
$\bR_2$ contains an element $\approx \ \in \bR_2$, called the identity. \\
\ (iv) The set of logic symbols $\Rightarrow$, $\FF$, and
$\forall$.}

Let $T(\mL)$ be the set of terms which is defined inductively as the
smallest set such that \\ \ (i) $\{x_1, x_2, ...\} \subset T(\mL)$.
\\ \ (ii) if $t_1, t_2, ..., t_n$ are terms and $f \in F_n$ then
$f(t_1, t_2, ..., t_n)$ is a term.

Let $F_a(\mL)$ be the set of expressions (called atomic formulas)
$r(t_1, t_2, ..., t_n)$ where $r \in \bR_n$ and $t_1, t_2, ..., t_n
\in T(\mL)$ (thus $\approx(t_1, t_2) \in F_a(\mL)$, which will be
denoted as $t_1 \approx t_2$).

Let $F(\mL)$ be the set which is defined inductively as the smallest
set such that \\ \ (i)  $F_a(\mL) \subset F(\mL)$. \\ \ (ii)  $\FF
\in F(\mL)$. \\ \ (iii) if $p, q \in F(\mL)$  then $p \Rightarrow q
\in F(\mL)$. \\ \ (iv) If $p \in F(\mL)$ then $\forall p \in
F(\mL)$.
\\ \ An element in $F(\mL)$ is called a \la{formula} of $\mL$.

We define a multiplication $T(\mL) \times T(\mL)^{\mN} \ar T(\mL)$
inductively:
\\ \ a. $x_i[t_1, t_2, ...] = t_i$ for any $t_1, t_2, ... \in T(\mL)$.
\\ \ b. $f(s_1, s_2, ..., s_n)[t_1, t_2, ...] = f(s_1[t_1, t_2, ...],
s_2[t_1, t_2, ...], ..., s_n[t_1, t_2, ...])$. \\$T(\mL)$ is a clone
with the unit $[x_1, x_2, ...]$.

Next we define a multiplication $F(\mL) \times T(\mL)^{\mN} \ar
F(\mL)$ inductively: \\ \ (i) $r(p_1, p_2, ..., p_n)[t_1, t_2, ...]
= r(p_1[t_1, t_2, ...], p_2[t_1, t_2, ...], ...)$. \\ \ (ii)
$\FF[t_1, t_2, ...] = \FF$. \\ \ (iii) $(p \Rightarrow q) [t_1, t_2,
...] = (p[t_1, t_2, ...])\Rightarrow p[t_1, t_2, ...])$. \\ \ (iv)
$(\forall p)[t_1, t_2, ...] = \forall (p[x_1, t_1^+, t_2^+, ...]$,
where $t_i^+ = t_i[x_2, x_3, ...]$. \\Then $F(\mL)$ is a right
$T(\mL)^{\mN}$-act. So $F(\mL)$ is a locally finitary right
$T(\mL)$-algebra.

\df{Let $\mL$ be a first order language. An interpretation (or a
model) $M$ of $\mL$ consists of the following ingredients: \\ \ (i)
A non-empty set $D$, called the domain of the interpretation. \\ \
(ii)  For each function symbol $f \in \bF_n$ an assignment of an
$n$-place operation $f^M$ in $D$, i.e. a function from $D^n$ to $D$.
\\ \ (iii) For each predicate symbol $r \in \bR_n$ an assignment
of an $n$-place relation $r^M$ in $D$, i.e. a subset of $D^n$. We
assume $\approx^M = \{(d, d)|d \in D\}$.  }

Suppose $M$ is an interpretation of $\mL$.\\
We first define a multiplication $T(\mL) \times D^{\mN} \ar D$
inductively: \\ \ a. $x_i[d_1, d_2, ...] = d_i$ for any $d_1, d_2,
... \in D^{\mN}$. \\ \  b. $f(t_1, t_2, ..., t_n)[d_1, d_2, ...] =
f^M(t_1[d_1, d_2, ...], t_2[d_1, d_2, ...], ..., t_n[d_1, d_2,
...])$. \\Then $D$ is a left $T(\mL)$-algebra.

Let $2 = \{0, 1\}$. We define a multiplication $F(\mL) \times
D^{\mN} \ar 2$ inductively:
\\ \ a. $r(t_1, t_2, ..., t_n)\td = 1$ if and only if
$(t_1\td, t_2\td, ..., t_n\td) \in r^M$ for any $\td \in D^{\mN}$ \\
\ b.  $\FF\td = 0$. \\ \  c.   $(p \Rightarrow q) \td = 1$ if and
only if $p\td = 0$  or $q\td = 1$.
\\ \ d.  $(\forall p)[d_1, d_2, ...] = 1$ if and only if $p(d, d_1,
d_2, ...) = 1$ for any $d \in D$.

\df{Let $M$ be an interpretation of $\mL$.
\\ \ 1. A sequence
$(d_1, d_2, ...)$ in $D^{\mN}$  satisfies a formula $p$ iff $p[d_1,
d_2, ...] = 1$. \\ \ 2. A formula $p$ is true for the interpretation
$M$ (written $\models_M p$) iff every sequence in $D^{\mN}$
satisfies $p$. \\ \ 3. $p$ is said to be
false for $M$ iff no sequence in $D^{\mN}$ satisfies $p$. \\
4. $M$ is said to be a model for a set $\Gamma$ of formulas if and
only if every $p$ in $\Gamma$ is true for $M$.}

\df{1. A formula $p$ is said to be logically valid iff $p$ is true
for every interpretation of $\mL$.
\\ \ 2. $p$ is said to be logically imply $q$ iff in
every interpretation, every sequence that satisfies $p$ also
satisfies $q$.
\\ \ 3. $p$ is said to be a logical consequence  of a set $\Gamma$ of formulas iff
 for every interpretation, every sequence that satisfies every
formula in $\Gamma$ also satisfies $p$. }

If $p \in F(\mL)$ and $a \in T(\mL)$ let $p[a/x_i] = p[x_1, x_2,
..., x_{i-1}, a, x_{i+1}, ...]$. In the following we assume $y, z,
w, ... \in \{x_1, x_2, ...\}$, which are called \la{(syntactical)
variables}. We say a variable $y$ is free for $p \in F(\mL)$ if $a
\ne a[y^+/y]$; otherwise we say that $p$ is independent of $y$.
Denote by $FV(p)$ the set of free variables of $p$, which is always
a finite set. If  $FV(p) = \emptyset$ the we say that $p$ is a
sentence (or that $p$ is closed, or $p$ has rank $0$). We say a
formula $p$ has a rank $n > 0$ if $p[[x_1, x_2, .., x_n]] = p$. If
$p$ has hank $n \ge 0$ then $FV(p) \subseteq \{x_1, x_2, ...,
x_n\}$.

Suppose $\mL$ is any first-order language. We introduce some further
operations: $ y \approx z = \approx[y, z, x_3, x_4, ...] $, $\exists
p = \neg \forall \neg p$.

For any variable $x_i$ we derive two maps $\lf x_i, \lx x_i: P \ar
P$ by
\[\lf x_i.p = \forall (p[x_2, x_3, ..., x_{i-1}, x_i, x_1, x_{i+2}, ...]\]
\[\lx x_i.p = \exists (p[x_2, x_3, ..., x_{i-1}, x_i, x_1, x_{i+2}, ...]\]
 Note that $\lf x_1.p = (\forall p)[x_2, x_3, ...]$. So we
have $\forall p = (\lf x_1.p)[x_1, x_1, x_2, ....]$, and \[\lf x_i.p
= (\lf x_1.(p[x_2, x_3, ..., x_{i-1}, x_i, x_1, x_{i+2}, ...]))[x_1,
x_1, x_2, ...].\] hence $\forall$ and $\lf x_1$ determines each
other, and each $\lf x_i$ can be derived from $\lf x_1$.

\ex{Suppose $M$ is an interpretation for $\mL$. For any formula $p$
we have  $(\lf x_i. p)\td = \forall (p[x_2, x_3, ..., x_i, x_1,
x_{i+2}, ...])\td = 1$ for any sequence $\td$ if and only if
\[(p[x_2, x_3, ..., x_i, x_1, x_{i+2}, ...])[d, d_1, d_2, ...] =
p([d_1, d_2, ..., d_{i-1}, d, d_{i+1}, ...] = 1\] for all $d \in D$,
which is the classical definition for $\lf x_i. p$.}

\lm{1. If $p$ has rank $n > 0$ then $\forall p$ has rank $n-1$. \\ \
2. If $p$ is a sentence then $\forall p $ and $\lf x_i.p$ are
sentences.
\\ \ 3. If $p$ has rank $n > 0$ then $\forall^n p$ is a sentence.
\\ \ 4. $\lf x_i.p$ is independent of $x_i$.
\\ \ 5. If $p$ is independent of $y$ then $ \lf z.b = \lf y. (b[y/z])$.
\\ \ 6.. If $p$ is independent of $y$ then  $\lf y.b = \forall (b^+)$.
\\ \ 7. If $y \ne z$ and $a \in T(\mL)$ is independent of $z$ then $(\lf z.b)[a/y] =
\lf z.(b[a/y])$. }

Now it is easy to translate the classical treatment of first order
theory into our setting and prove all the fundamental theorems of
first order theory.

\section{Predicate Algebras}\label{sec:pa}
\df{Let $\A$ be a clone. A predictive algebra with terms in $\A$ is
a right $\A$-algebra $P$ together with three homomorphisms of right
$\A$-algebras: \\ 1. $\Rightarrow: P^2 \ar P$. \\ 2. $\FF: P^0 \ar
P$ (i.e. $\FF$ is an element of $P$ of rank $0$.) \\ 3. $\forall:
P^\A \ar P$.
\\We also assume that $P$ \la{has identity}, which is an element
$\approx \ \in P$ of rank $2$. }

Thus a  predicate algebra with terms in $\A$ is an $(\A,
\tau_p)$-algebra with a homomorphism $\forall: P^\A \ar P$ and an
element $\approx \ \in P$ of rank $2$. Any proposition algebra may
be viewed trivially as a predicate algebra with terms in $\A$ such
that all elements are closed and $\approx = \TT$.

\ex{Let $D$ be a left $\A$-algebra. Since $2 = \{0, 1\}$ is a
proposition algebra, $2^{D^{\mN}}$ is an $(\A, \tau_p)$-algebra. For
$p \in 2^{D^{\mN}}$ let $\forall p \in 2^{D^{\mN}}$ such that
$\forall p[d_1, d_2, ...] = 1$ if and only if $p[d, d_1, d_2, ...] =
1$ for all $d \in D$. Define $\approx \ \in 2^{D^{\mN}}$ such that
$\approx [d_1, d_2, ...] = 1$ if and only if $d_1 = d_2$. Then
$2^{D^{\mN}}$ is a predicate algebra with terms in $\A$, called the
\la{predicate set algebra for $D$}. }

\ex{Let $B$ be a right $\A$-algebra. Let $P_B$ be the set which is
defined inductively as the smallest set such that (a) $B \subset
P_B$. (b) $\FF \in P_B$. (c) if $p, q \in P_B$  then $p \Rightarrow
q \in P_B$ and $\forall p \in P_B$. Then $P_B$ is a predicate
algebra with terms in $A$, called the \la{free predicate algebra
over $B$}.  In particular, if $S$ is any set then $S \times \A$ is
the free right $\A$-algebra over $S$, and $P_{S \times \A}$ is the
\la{free predicate algebra over $S$}. Note that if $B$ is locally
finitary then so is $P_B$.}

\ex{The initial predicate algebra with terms in a clone $\A$ is a
locally finitary predicate algebra $Eq(\A)$, called the
\la{equational logic} for $\A$.}

Let $P$ be a predicate algebra with terms in $\A$. For any variable
$x_i$ we introduce a new map $\lf x_i: P \ar P$ by  $\lf x_i.p =
\forall (p[x_2, x_3, ..., x_{i-1}, x_i, x_1, x_{i+2}, ...]$. Note
that $\lf x_1.p = (\forall p)^+.$ So we have $\forall p = (\lf
x_1.p)^-$, and \[\lf x_i.p = (\lf x_1.(p[x_2, x_3, ..., x_{i-1},
x_i, x_1, x_{i+2}, ...])^-.\] Since $\forall$ and $\lf x_i$ can be
derived from $\lf x_1$, a predicate algebra can also be defined
using $\lf x_1$ (instead of $\forall)$ as the basic operation. Let
$\exists p = \neg \forall \neg p$ and $\lx x_i. p = \neg \lf x_i.
\neg p$.

\ex{Suppose $D$ is a left $\A$-algebra. For any $p \in 2^{D^{\mN}}$
and any $x_i$ we have $(\lf x_i. p)\td = \forall (p[x_2, x_3, ...,
x_i, x_1, x_{i+2}, ...])\td = 1$ for any $\td \in D^{\mN}$ if and
only if $p[x_2, x_3, ..., x_i, x_1, x_{i+2}, ...][d, d_1, d_2, ...]
= p([d_1, d_2, ..., d_{i-1}, d, d_{i+1}, ...]) = 1$ for all $d \in
D$.}

An \la{interpretation of $P$} (or a \la{P-structure}, or a \la{model
of $P$}) is a pair $(D, \mu)$ consisting of a left $\A$-algebra $D$
and a homomorphism $\mu: P \ar P(D^{\mN})$ of predicate algebras. We
say $p \in P$ is \la{logical valid} (written $\models p$) if for any
interpretation $(D, \mu)$ we have $\mu(p) = \TT$. If $p, q \in P$
then we say that $p$ and $q$ are \la{logically equivalent} (written
$p \equiv q$) if $p \Leftrightarrow q$ is logically valid. The
relation $\equiv$ is a congruence relation on $P$. The set of
congruence classes of $P$ with respect to $\equiv$ is a predictive
algebra called the \la{Lindenbaum-Tarski algebra of $P$}, denote by
$LT(P)$.

\df{A predicate algebra with terms in a clone $\A$ is called a
\la{quantifier algebra with terms in $\A$} if $p \equiv q$ implies
that $p = q$, i.e. $P = LT(P)$.}

Any predicate set algebra is a quantifier algebra. By definition the
class of quantifier algebras is the variety generated by all
predicate set algebras. A quantifier algebra is a Boolean algebra
with respect to the operations $\vee, \wedge, \neg$ such that each
$a \in \A$ induces an endomorphism of this Boolean algebra. We also
have existential quantifiers $(\lx x_1), (\lx x_2), ...,$. These
data determined a polyadic algebra with terms in $\A$.

\te{(cf.~\cite{pinter:firstorder}) A locally finitary predicate
algebra $P$ with terms in a clone $\A$ is a quantifier algebra if
and only if the following conditions are satisfied for all $a, b \in
P$ and variables $y, z$:
\\1. $P$ is a Boolean algebra with respect to $\vee, \wedge, \neg, \FF, \TT$.
\\2. $\exists (a \vee b) = \exists a \vee \exists b$.
\\3. $a \leq (\exists a)^+$.
\\4. $x_1 \approx x_1 = \TT$.
\\5. $a \wedge (y \approx z) \leq a[z/y]$, where $a[z/y] = a[x_1, x_2, ..., y^-, z, y^+, ...]$.}

\ex{Let $\mL$ be a first-order language. Then
\\1. $T(\mL)$ is a locally finitary clone.
\\2. $F(\mL)$ is a locally finitary predicate algebra with terms
in  $T(\mL)$.
\\3. If $M$ is an interpretation of $\mL$ then the multiplication
$T(\mL) \times D^{\mN} \ar D$ turns $D$ into a left
$T(\mL)$-algebra, and the multiplication $F(\mL) \times D^{\mN} \ar
2$ induces a homomorphism of predicate algebras $F(\mL) \ar
2^{D^{\mN}}$. \\ 4. Conversely if $D$ is a left $T(\mL)$-algebra
then any homomorphism $\mu: F(\mL) \ar 2^{D^{\mN}}$ induces an
interpretation $M$ for $\mL$ such that $f^M = f(x_1, x_2, ..., x_n)
\in D$ for any $f \in  \bF_n$ and for any $r \in \bR_n$ we have
$(d_1, d_2, ..., d_N) \in r^M$ iff \[\mu(r(x_1, x_2, ..., x_n))(d_1,
d_2, ..., d_n) = 1.\]}

\end{document}